\documentclass{article}
\usepackage{spconf,amsmath,graphicx}
\usepackage{comment}
\usepackage{cite}
\usepackage{multirow}
\usepackage{amsfonts}
\usepackage{booktabs}

\usepackage{hyperref}
\title{LSZone: A Lightweight Spatial Information Modeling Architecture for Real-time In-car Multi-zone Speech Separation}
\name{
\begin{tabular}{@{}c@{}}
Jun Chen$^{1,2}$, Shichao Hu$^{1}$, Jiuxin Lin$^{1,2}$, Wenjie Li$^1$, Zihan Zhang$^{1,3}$, Xingchen Li$^3$,\\
\textit{JinJiang Liu$^{2}$, Longshuai Xiao$^{1}$, Chao Weng$^{1}$, Lei Xie$^3$, Zhiyong Wu$^{2}$}
\end{tabular}
\vspace{-0.1cm}
}
\address{%
$^1$ Huawei Technologies Co., Ltd., Shanghai, China\\
$^2$ Shenzhen International Graduate School, Tsinghua University, Shenzhen, China\\
$^3$ School of Software, Northwestern Polytechnical University, Xi\'an, China\\
\small y-chen21@mails.tsinghua.edu.cn,\; xiaolongshuai@huawei.com
}
\begin{document}
\ninept
\maketitle
\begin{abstract}
\vspace{-0.05cm}
In-car multi-zone speech separation, which captures voices from different speech zones, plays a crucial role in human-vehicle interaction.
Although previous SpatialNet has achieved notable results, its high computational cost still hinders real-time applications in vehicles.
To this end, this paper proposes LSZone, a lightweight spatial information modeling architecture for real-time in-car multi-zone speech separation.  
We design a spatial information extraction-compression (SpaIEC) module that combines Mel spectrogram and Interaural Phase Difference (IPD) to reduce computational burden while maintaining performance.
Additionally, to efficiently model spatial information, we introduce an extremely lightweight Conv-GRU crossband-narrowband processing (CNP) module.
Experimental results demonstrate that LSZone, with a complexity of 0.56G MACs and a real-time factor (RTF) of 0.37, delivers impressive performance in complex noise and multi-speaker scenarios.
\end{abstract}
\vspace{-0.1cm}

\vspace{-0.35cm}
\section{Introduction}
\vspace{-0.15cm}
\label{sec:intro}

Recent advances in deep learning technologies, including Automatic Speech Recognition (ASR) \cite{gulati2020conformer, xu2023cb, shi2024advancing}, Speaker Verification (SV) \cite{li2024safeear, zeng2024spoofing}, and Text-to-Speech (TTS)\cite{ren2019fastspeech, anastassiou2024seed}, have significantly enhanced human-machine interactions. 
Among them, human-vehicle interaction \cite{capallera2022human} has rapidly developed into a mainstream trend, driving the need for more natural and convenient interaction methods. 
This has led to the rise of in-car multi-zone speech separation, which utilizes distributed microphone arrays to precisely capture voices from different zones, enabling more accurate and efficient speech recognition and interaction. 
Nevertheless, compared to traditional speech separation tasks, the in-car multi-zone speech separation faces additional challenges.
The driving environment often involves low Signal-to-Noise Ratio (SNR) signal and multiple simultaneous speakers, creating complex acoustic scenarios. 
Additionally, the distributed microphone arrays require the model to process more audio channels, while being constrained by limited computational resources and strict low-latency requirements. 
Therefore, the research into in-car multi-zone speech separation is of paramount importance and presents considerable challenges.

Early approaches \cite{hansen2010analysis, vu2010small} to in-car multi-zone speech separation relied on traditional beamforming based on matrix operations, yet their separation performance remained inherently limited.
With the emergence of deep learning, neural beamforming techniques \cite{heymann2016neural, erdogan2016improved} have been developed to overcome these constraints.
Notably, Zoneformer \cite{xu2023zoneformer} integrates sub-band processing with full-frequency information, leveraging an RNN-Transformer-based neural beamforming architecture to enable efficient in-car multi-zone speech separation.
Meanwhile, DualSep \cite{wang2024dualsep} adopts a different hybrid approach that incorporates digital signal processing and neural networks.
By leveraging fixed beamforming to reduce computational overhead, employing independent vector analysis (IVA) for spatial priors, and utilizing a dual-encoder mechanism to distinctly capture spatial and spectral features, DualSep strikes an optimal balance between efficiency and performance, making it a lightweight and effective solution for real-time in-car multi-zone speech separation.

%Nonetheless, compared to the previously discussed 
Compared to
RNN-Transformer-based temporal modeling architectures \cite{xu2023zoneformer} and UNet structures \cite{wang2024dualsep}, recent breakthroughs in time-frequency domain alternating modeling approaches, such as TF-GridNet \cite{wang2023tf} and BSRNN \cite{luo2023music}, have demonstrated superior efficiency and effectiveness in the field of single-channel speech enhancement and separation.
SpatialNet \cite{quan2024spatialnet} takes this a step further by incorporating spatial information into similar architectures, alternating between Crossband and Narrowband modules to model spatial, frequency, and temporal information. 
This has led to outstanding results in multi-channel speech separation and enhancement tasks. 
However, when SpatialNet is applied to in-car multi-zone speech separation, higher sampling rates increase feature dimensionality, while more distributed microphones and speech zones lead to a rise in input-output channels. 
These factors collectively result in a surge in computational load, making real-time processing unfeasible with the car’s limited computational resources.
% Adjusting hyperparameters alone is insufficient to resolve this challenge, as it fails to provide an effective balance between performance and computational demands.

In order to address the aforementioned challenges, this paper introduces LSZone, a lightweight spatial information modeling framework for real-time in-car multi-zone speech separation.
First of all, we design a SpaIEC module that reduces feature dimensionality by integrating Mel spectrogram with IPD.
This module not only preserves performance but also significantly reduces computational load.
Furthermore, to optimize spatial information capture, we introduce an ultra-lightweight Conv-GRU CNP module
, which alternates convolutional crossband modeling and GRU-based narrowband modeling to efficiently integrate spatial, frequency, and temporal cues with minimal computational overhead.
Experimental results demonstrate that LSZone delivers remarkable computational efficiency, with a computational cost of 0.56G MACs and a RTF of 0.37, alongside superior performance in challenging noisy and multi-speaker environments.

\begin{figure*}[!htbp]
	\centering
% 	\vspace{-0.1cm}
    \includegraphics[width=1.0\linewidth]{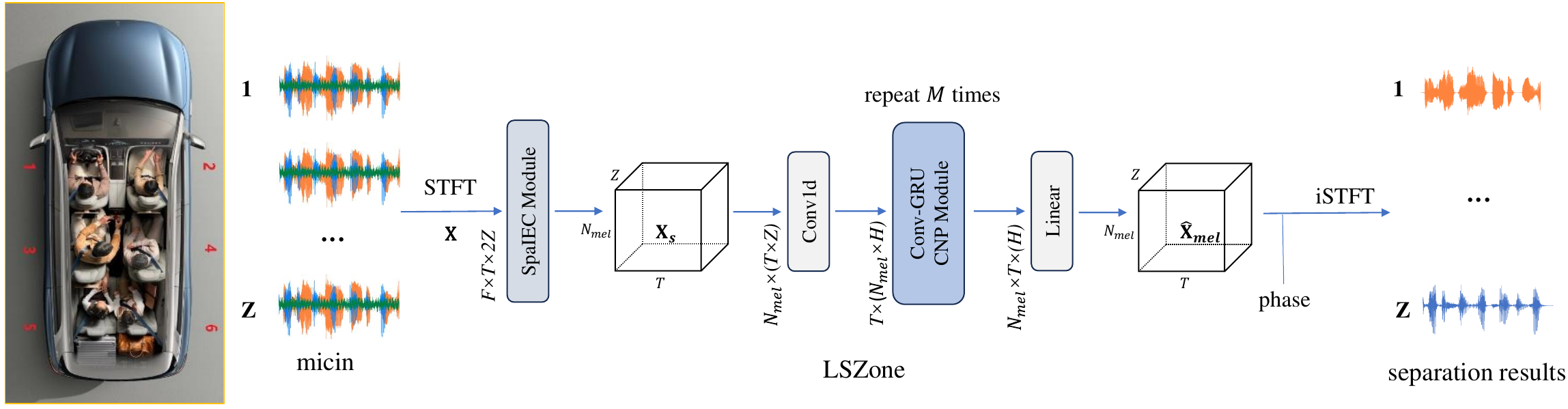}
    \vspace{-0.8cm}
	\caption{The overall diagram of LSZone. It mainly comprises the SpaIEC Module and a backbone structure consisting of Conv1D, Conv-GRU CNP Modules, and a linear layer.}
% 	\caption{The details of the SubInter module. The ``Hidden" indicates the hidden representations. The ``Global" denotes the output of the second linear layer with cross-band global information.}
	\label{fig:total_arch}
	 \vspace{-0.4cm}
\end{figure*}

\begin{figure*}[!htbp]
	\centering
% 	\vspace{-0.1cm}
    \includegraphics[width=0.68\linewidth]{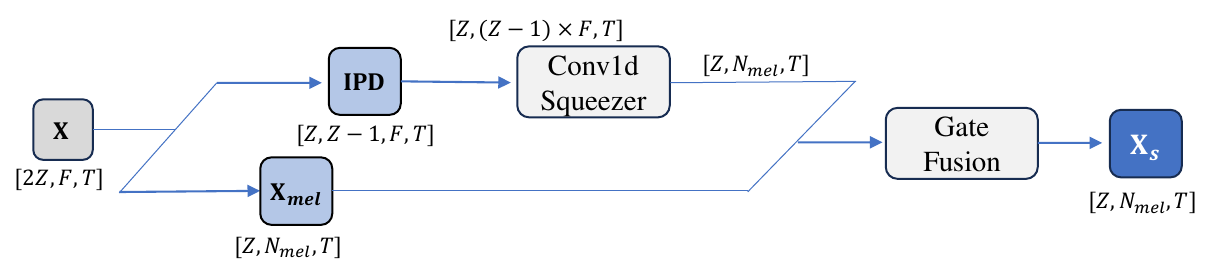}
    \vspace{-0.4cm}
	\caption{The details of the SpaIEC Module, which primarily consists of a Conv1D Squeezer and a Gate Fusion mechanism.}
% 	\caption{The details of the SubInter module. The ``Hidden" indicates the hidden representations. The ``Global" denotes the output of the second linear layer with cross-band global information.}
	\label{fig:spaiec}
	 \vspace{-0.4cm}
\end{figure*}

\vspace{-0.2cm}
\section{Methodology}
\label{sec:method}
\vspace{-0.15cm}
For in-car scenario, assume there are $Z$ zones, each with one corresponding microphone, and $P$ speakers $(P \le Z)$ within those zones. 
The input audio $y(n)$ can be defined as follows:
\vspace{-0.1cm}
\begin{equation}
y(n) = \sum_{i=0}^{P-1}s_i(n) + \eta(n),
\end{equation}
where $s_i(n)$ and $\eta(n)$ represent the audio of the $i$-th speaker and the noise, respectively, after convolution with the room impulse responses (RIRs) for $Z$ channels.
If the $i$-th speaker is located in the $j$-th zone $(0 \le j < Z)$, we denote the channel of $s_i(n)$ that corresponds to the $j$-th zone as $s_{i, j}(n)$.
The objective of this work is to suppress the noise from the input audio and separate the mixed speech into its corresponding channels, yielding $\hat{s}(n)$:
\vspace{-0.1cm}
\begin{equation}
\hat{s}(n) = \sum_{i=0}^{P-1}s_{i,j}(n).
\end{equation}
\vspace{-0.1cm}

Towards this goal, we propose a lightweight spatial information modeling architecture called LSZone.
The overall diagram of LSZone is illustrated in Figure \ref{fig:total_arch}. 
The model takes the multi-channel spectrogram $\mathbf{X} \in \mathbb{R}^{F \times T \times 2Z}$, formed by concatenating the real and imaginary parts obtained through Short-Time Fourier Transform (STFT) applied to the multi-channel audio, as its input
, where $F$ and $T$ denote the number of frequency bins and time frames, respectively.
Subsequently, the SpaIEC module compresses and extracts features from the input, generating $\mathbf{X}_s \in \mathbb{R}^{N_{mel} \times T \times Z}$
, where $N_{mel}$ is the number of Mel filter banks.
The $\mathbf{X}_s$ is processed through the core network, which includes a Conv1D layer, $M$ stacked Conv-GRU CNP modules, and a linear layer, to capture spatial, frequency, and temporal information, yielding the predicted Mel spectrogram $\mathbf{\hat{X}}_{mel} \in \mathbb{R}^{N_{mel} \times T \times Z}$.
Finally, the $\mathbf{\hat{X}}_{mel}$ is integrated with the phase information of the original audio and passed through the inverse Short-Time Fourier Transform (iSTFT) to obtain the separation result.
As in \cite{quan2024spatialnet}, we train the model by calculating the Mean Squared Error (MSE) loss between the multi-channel complex spectrograms of the separated results and the corresponding ground truth.

\vspace{-0.05cm}
\subsection{SpaIEC Module}

One of the primary reasons for SpatialNet’s high computational complexity is the large feature dimensionality, especially in the frequency domain $F$, where both real and imaginary parts of the spectrogram are processed.
In contrast, most modern ASR systems, whether based on large language models (LLMs) \cite{bai2024seed, an2024funaudiollm, chu2024qwen2} or traditional attention-based encoder-decoder (AED) architectures \cite{gulati2020conformer,  xu2024hydraformer}, take Mel spectrograms or Mel-frequency cepstral coefficients (MFCCs) as input features. 
Drawing inspiration from this, we believe that focusing on Mel spectrogram denoising and multi-zone speech separation is sufficient to maintain the performance of ASR systems in human-vehicle interactions, while effectively reducing the model's feature dimensions. 
Nonetheless, since Mel spectrograms lack phase information, relying solely on them could potentially discard some valuable spatial information. 
To mitigate this, we introduce the SpaIEC module, which integrates Mel spectrograms with spatial information to effectively lower the feature dimensionality processed by the model.

Specifically, we adopt the IPD as a spatial feature to supplement the spatial information. 
As a directional feature, IPD aids in capturing the spatial location of sound sources in multi-microphone arrays \cite{gu2024rezero}.
For each channel $z$, we compute its $\mathbf{IPD}_z \in \mathbb{R}^{(Z-1) \times F \times T}$ with the other $Z-1$ channels as follows:
\begin{align*}
\mathbf{IPD}_{z,k} &= \mathbf{X}_{phase}[z] - \mathbf{X}_{phase}[k], \quad k \neq z, \\
\mathbf{IPD}_z &= \bigoplus_{k \neq z} \mathbf{IPD}_{z,k},
\end{align*}
where $\bigoplus$ denotes stacking, and $\mathbf{X}_{phase}$ represents the phase extracted from the input $\textbf{X}$. We stack the IPD features from all $Z$ channels to obtain the $\mathbf{IPD} \in \mathbb{R}^{Z \times (Z-1) \times F \times T}$ for the entire multi-channel audio:
\vspace{-0.05cm}
$$
\mathbf{IPD} = \bigoplus_{z=0}^{Z-1} \mathbf{IPD}_{z}.
$$
As shown in Figure \ref{fig:spaiec}, after processing the input $\mathbf{X}$, we obtain both the Mel spectrogram $\mathbf{X}_{mel}$ and the $\mathbf{IPD}$. 
The $\mathbf{IPD}$ is then compressed and extracted through a Conv1D Squeezer, aligning with the feature dimensions of $X_{mel}$. 
Subsequently, a Gate Fusion module, composed of a Conv1D layer and a Sigmoid activation function, performs an interactive fusion of these two, resulting in the reduced-dimensional feature $\mathbf{X}_s$ that encapsulates both Mel and spatial information.

\begin{figure}[t]
	\centering
% 	\vspace{-0.1cm}
    \includegraphics[width=0.97\linewidth]{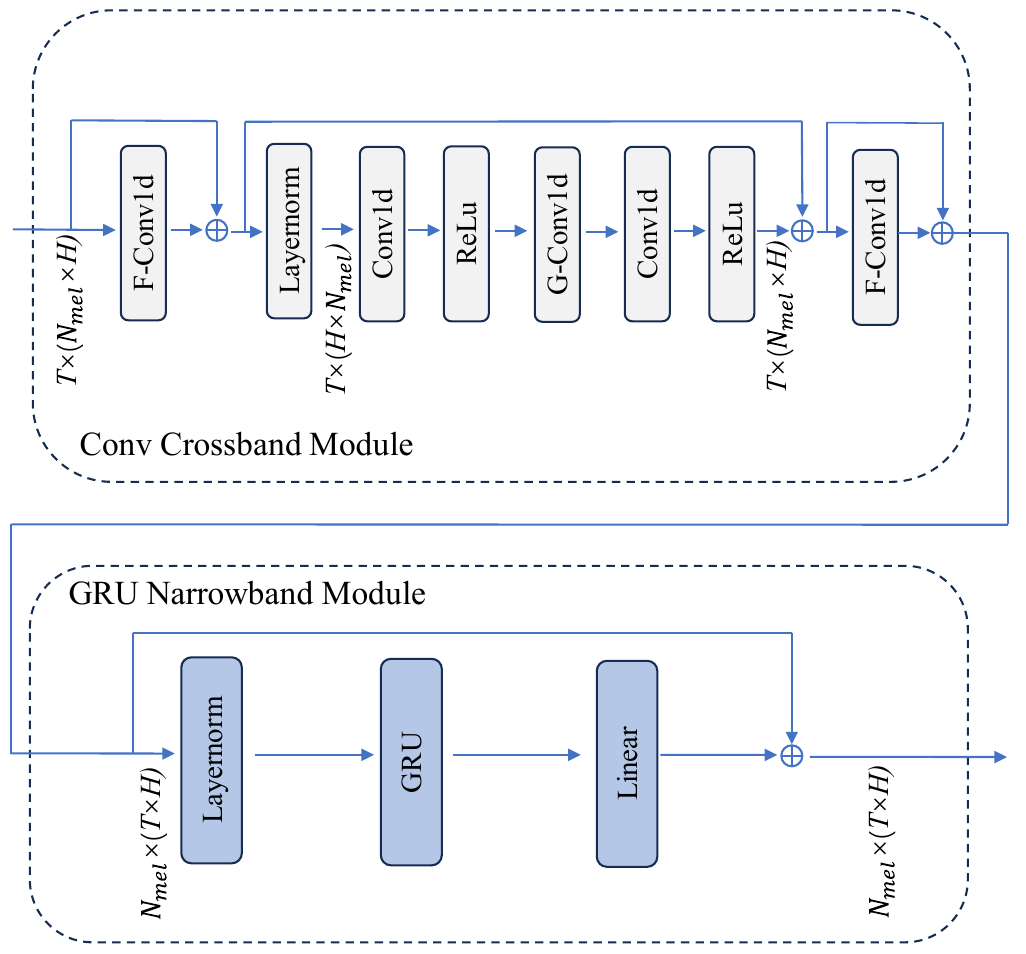}
    \vspace{-0.3cm}
	\caption{The specifics of the Conv-GRU CNP Module. The ``F-Conv1d" indicates the Frequency Conv1d while the ``G-Conv1d" denotes the Group Conv1d.}
% 	\caption{The details of the SubInter module. The ``Hidden" indicates the hidden representations. The ``Global" denotes the output of the second linear layer with cross-band global information.}
	\label{fig:cnp}
	\vspace{-0.4cm}
\end{figure}

\vspace{-0.15cm}
\subsection{Conv-GRU CNP Module}
In the field of source separation, architectures such as BSRNN and SpatialNet are built around alternating time-frequency domain modeling structures.
Particularly, BSRNN employs LSTMs to model both the time and frequency domains \cite{luo2023music}, but the non-parallel processing in the frequency domain results in significant time consumption during single-frame streaming inference.
On the other hand, SpatialNet utilizes 1D convolutions and linear layers in the Crossband block, and Attention along with time-domain convolutions in the Narrowband block \cite{quan2024spatialnet}, leading to substantial computational overhead. 
In this work, to address the challenges posed by these alternating time-frequency domain structures, we propse an extremely lightweight Conv-GRU CNP module.

The architecture of the Conv-GRU CNP module is depicted in Figure \ref{fig:cnp}. 
It is primarily comprised of the Conv Crossband module, which models the spatial-frequency domain, and the GRU Narrowband module, which focuses on the spatial-temporal domain.
The Conv Crossband module leverages parallelized operations such as Frequency Conv1D, Conv1D and Group Conv1D to efficiently process spatial information, represented by $H$, and frequency-domain features, denoted by $N_{mel}$.
Meanwhile, the GRU Narrowband module, composed of Layernorm, GRU, and Linear layers, performs effective processing of spatial information $H$ and temporal information $T$ with minimal computational overhead. 
By alternating between these two modules, the Conv-GRU CNP module achieves efficient interactive modeling across the spatial, frequency, and temporal domains.

\vspace{-0.3cm}
\section{Experimental Setup}
\vspace{-0.1cm}
\subsection{Datasets and Evaluation Metrics}
\textbf{Datasets}: We design the task as a typical in-car 6-zone speech separation, with each speech zone equipped with a corresponding microphone. 
In accordance with \cite{xu2023zoneformer, wang2024dualsep}, we employ the AISHELL-1 \cite{bu2017aishell} for clean speech dataset and the DNS Challenge 5 \cite{dubey2023icassp} for noise dataset, generating the multi-speaker in-car dataset via RIRs.
A total of 50k RIRs are generated through the image method, simulating a vehicle cabin with dimensions of 1.45m in width, 2.70m in length, and 1.25m in height, and an RT60 range between [50ms, 90ms].
These RIRs are randomly selected and combined with the clean speech and noise to simulate the in-car conditions.
One to six passengers are randomly assigned to the six zones and speak simultaneously. 
The voices of different passengers are mixed at a signal-to-interference ratio (SIR) of [-6, 6] dB, followed by mixing speech with noise at a SNR of [-10, 20] dB. 
In this manner, we create a training set of 120k clips and a validation set of 5k clips. 
Additionally, a test set of 1.2k clips is constructed in the same manner, with subsets for single-speaker, two-speaker and three-speaker scenarios, each containing 400 clips.
To maintain evaluation integrity, the noise and RIRs employed across the training, validation, and test sets are mutually exclusive.

\textbf{Evaluation Metrics}: We evaluate the model across two fundamental axes: computational load and performance effectiveness.
For computational load, we assess the number of parameters, computational complexity, and RTF on a single Intel Xeon Platinum 8180 (2.50GHz) CPU. 
In terms of performance effectiveness, given that in-car multi-zone speech separation is predominantly utilized in human-vehicle interaction scenarios, we employ SenseVoice\footnote{\href{https://github.com/FunAudioLLM/SenseVoice}{https://github.com/FunAudioLLM/SenseVoice}} as our ASR system and use the character error rate (CER) and false intrusion rate (FIR) derived from its recognition results as evaluation metrics.
The FIR is a metric that we design to evaluate errors in human-vehicle interactions caused by audio leakage into incorrect speech zones, a scenario that must be minimized.
Specifically, we gather recognition results for voice clips from non-speaking zones and define the FIR as the ratio of clips with non-empty recognition results to the total number of clips. 
A lower FIR reflects better suppression of audio leakage into incorrect zones.

\subsection{Training setup and baselines}
We adopt an audio sampling rate of 16 kHz and apply the STFT using a Hanning window with a window length of 32 ms and a frame shift of 16 ms.
During training, input-target pairs are generated as fixed-length sequences, with each sequence set to a length of 3s.
The Adam optimizer is employed, with the learning rate initialized at 1e-3 and decayed exponentially according to $lr = 0.001\times 0.99^{epoch}$.

In this study, we compare our proposed LSZone with several competitive baselines, including Zoneformer, DualSep and SpatialNet.
It is important to note that all methods are causal.
(1) \textbf{Zoneformer \& DualSep}: For both of these models, aside from adjustments to the input-output channels, all other hyperparameters are kept consistent with those in the original works.
(2) \textbf{SpatialNet}: We use the causal version of SpatialNet, with the hidden feature set to 72, the hidden units set to 144, and both the crossband and narrowband blocks stacked and repeated 4 times.
(3) \textbf{LSZone}: For LSZone, we set $N_{mel}=64$. The model consists of 6 repeated Conv-GRU CNP modules, with the hidden feature set to 72 and the hidden units set to 144.
\vspace{-0.3cm}

\begin{table*}[!htbp]
    \begin{center}
    \caption{Comparison with baselines on CER and FIR metrics for the overall test set and subsets (single, two, and three speakers) using the SenseVoice ASR system.} 
    % \vspace{-0.05cm}
    \label{table:large_total_compare}
    \scalebox{0.94}{
    \begin{tabular}{ccccccccccccc}
    \toprule
        \multirow{2}{*}{Method}  & \multirow{2}{*}{\shortstack{Param\\(M)}} & \multirow{2}{*}{\shortstack{MACs\\(G)}} & \multirow{2}{*}{RTF} &  \multicolumn{2}{c}{Overall} & \multicolumn{2}{c}{Single Speaker} & \multicolumn{2}{c}{Two Speakers} & \multicolumn{2}{c}{Three Speakers} \\
    \cmidrule(lr){5-6} \cmidrule(lr){7-8} \cmidrule(lr){9-10} \cmidrule(lr){11-12}
    &  &  &  & CER$\downarrow$ & FIR$\downarrow$ & CER$\downarrow$ & FIR$\downarrow$ & CER$\downarrow$ & FIR$\downarrow$ & CER$\downarrow$ & FIR$\downarrow$  \\
    \midrule
    Unprocessed & - & - & - & 37.97\% & 90.78\% & 12.36\% & 88.44\%  & 30.80\% & 91.62\% & 50.72\% & 93.17\% \\
    Ground Truth & - & - & - & 5.65\% & 0.00\% & 5.57\% & 0.00\%  & 5.68\% & 0.00\% & 5.65\% & 0.00\% \\
 
    \midrule
    Zoneformer & 1.85 & 1.13 & 1.35 & 30.03\% & 68.17\% & 12.60\% & 70.44\%  & 30.16\% & 67.50\% & 42.92\% & 65.67\% \\
    DualSep & \textbf{0.83} & 0.79 & 2.02 & 25.94\% & 69.26\% & 10.75\% & 62.00\%  & 20.67\% & 70.38\% & 34.21\% & 78.67\% \\
    SpatialNet & 0.89 & 6.98 & 2.91 & 17.59\% & 14.61\% & 8.42\% & 17.00\%  & 14.78\% & 11.50\% & 22.45\% & 15.17\% \\
    \midrule
    LSZone & 1.10 & \textbf{0.56} & \textbf{0.37} & \textbf{17.20\%} & \textbf{10.26\%} & \textbf{8.39\%} & \textbf{8.00\%}  & \textbf{13.61\%} & \textbf{9.50\%} & \textbf{22.35\%} & \textbf{14.67\%} \\
    \bottomrule
    \end{tabular}}
    \end{center}
    \vspace{-0.7cm}
\end{table*}

\begin{table}[t]
    \begin{center}
     \vspace{-0.3cm}
    \caption{CER and FIR performance in the study \ref{sec:spaiec} of SpaIEC module on the overall test set.}
    % \vspace{-0.05cm}
    \label{table:spaiec}
    \scalebox{0.85}{
    \begin{tabular}{lccccc}
    \toprule
    Model  & Param(M) & MACs(G) & RTF & CER$\downarrow$ & FIR$\downarrow$ \\
    \midrule

    Mel-LSZone   & 1.13 & 1.32 & 0.43 & 17.36\% & 12.39\%  \\    
    Phase-LSZone    & 1.10 & 0.56 & 0.36 & 18.70\% &  13.04\% \\
    LSZone    & 1.10 & 0.56 & 0.37 & \textbf{17.20\%} & \textbf{10.26\%} \\
    \bottomrule
    \end{tabular}}
    \end{center}
    \vspace{-0.55cm}
\end{table}

\begin{table}[t]
    \begin{center}
     \vspace{-0.2cm}
    \caption{CER and FIR performance in the study \ref{sec:cnp} of Conv-GRU CNP module on the overall test set.}
    % \vspace{-0.05cm}
    \label{table:cnp}
    \scalebox{0.85}{
    \begin{tabular}{lcccc}
    \toprule
    Model  &  MACs(G) & RTF & CER$\downarrow$ & FIR$\downarrow$ \\
    \midrule

    SpatialNet-o   &  6.98 & 2.91 & 17.59\% & 14.61\%  \\    
    SpatialNet-dLSTM    & 1.43 & 1.26 & 17.42\% &  14.91\% \\
    SpatialNet-Conv-GRU CNP    &  \textbf{1.38} & \textbf{0.93} & 17.38\% & 14.87\% \\
    \bottomrule
    \end{tabular}}
    \end{center}
    \vspace{-0.8cm}
\end{table}

\section{Results and Discussions}
\vspace{-0.1cm}
\subsection{Comparison with baselines}
Table \ref{table:large_total_compare} compares LSZone with the previously mentioned baselines.
As shown, compared to Zoneformer and DualSep, SpatialNet consistently reduces CER across all scenarios and achieves a even significant drop in the FIR metric, confirming the advantage of alternating time-frequency domain modeling with spatial information for the in-car multi-zone speech separation task.
Nevertheless, SpatialNet still suffers from high computational complexity and RTF.
In contrast, our proposed LSZone, with a low computational cost of 0.56G MACs and 0.37 RTF, outperforms Zoneformer, DualSep and SpatialNet in terms of both CER and FIR.
This demonstrates that LSZone is an efficient solution for in-car multi-zone speech separation.

\vspace{-0.1cm}
\subsection{Investigation of SpaIEC Module}
\label{sec:spaiec}
To assess the impact of the SpaIEC module, we compare LSZone with two variants: Mel-LSZone and Phase-LSZone, as shown in Figure \ref{fig:spaiec}.
In Mel-LSZone, the backbone model uses only the Mel spectrogram as input, while Phase-LSZone retains the SpaIEC structure but replaces the IPD with phase information.
For a fair comparison, we ensure that the number of parameters of each model are adjusted to similar scales.
The results in the table indicate that, with fewer parameters and computational resources, LSZone outperforms Mel-LSZone, especially in terms of FIR. 
This highlights the benefit of introducing IPD information via the SpaIEC module for in-car multi-zone speech separation, particularly in separating speech into the correct zones. 
Furthermore, the performance of Phase-LSZone is even worse than Mel-LSZone, which may be attributed to the model's inability to effectively extract the latent spatial information from phase alone with limited computational resources.
In contrast, introducing IPD, which carries prior knowledge, helps mitigate this issue.

\begin{table}[t]
    \begin{center}
     \vspace{-0.3cm}
    \caption{CER performance using SenseVoice and our AED architecture ASR systems on the overall test set.}
    % \vspace{-0.35cm}
    \label{table:asr}
    \scalebox{0.9}{
    \begin{tabular}{lcc}
    \toprule
    \multirow{2}{*}{Method}  &   \multirow{2}{*}{\shortstack{SenseVoice\\CER$\downarrow$}} & \multirow{2}{*}{\shortstack{AED architecture\\CER$\downarrow$}}\\
     &  & \\
    \midrule

    Zoneformer   & 30.03\% & 54.58\%   \\    
    DualSep   & 25.94\% & 47.12\%   \\
    SpatialNet   & 17.59\% & 35.54\%  \\
    LSZone   & \textbf{17.20\%} & \textbf{34.89\%}   \\
    \bottomrule
    \end{tabular}}
    \end{center}
    \vspace{-0.8cm}
\end{table}

\vspace{-0.1cm}
\subsection{Investigation of Conv-GRU CNP Module}
\label{sec:cnp}
We then further explore the role of the Conv-GRU CNP Module in Table \ref{table:cnp}.
We conduct exploratory experiments based on SpatialNet and propose several alternative configurations.
The model with the original SpatialNet structure is denoted as SpatialNet-o, while replacing the crossband and narrowband blocks in SpatialNet with frequency-domain and time-domain LSTMs from BSRNN results in SpatialNet-dLSTM. 
Eventually, replacing these blocks with the Conv-GRU CNP module gives rise to SpatialNet-Conv-GRU CNP.
All of these models have the hidden feature set to 72 and the hidden units set to 144, with the core modeling structure stacked and repeated 4 times.
The results indicate that SpatialNet-Conv-GRU CNP, with the lowest computational cost and RTF, achieves performance comparable to SpatialNet-o and SpatialNet-dLSTM in terms of CER and FIR.
This further validates that our proposed Conv-GRU CNP module provides a more lightweight and effective structure for the interactive modeling for spatial, frequency, and temporal information.

\vspace{-0.1cm}
\subsection{Model Generalization across ASR Systems}
\label{sec:asr}
Finally, we evaluate the model's generalization ability across different ASR systems in Table \ref{table:asr}.
In addition to SenseVoice, we also provide results from a smaller ASR system based on an AED structure \cite{gulati2020conformer}. 
The table shows that our proposed LSZone consistently yields the best CER performance, whether applied to SenseVoice or our smaller-scale AED-based ASR system.
This further supports the effectiveness of our improvements across a range of ASR backends.

\section{Conclusions}
\vspace{-0.1cm}
This paper proposes LSZone, an efficient framework designed for real-time in-car multi-zone speech separation.
By introducing the SpaIEC module, we effectively reduce feature dimensionality through the fusion of Mel spectrogram and IPD, striking a balance between preserving performance and minimizing computational overhead.
Moreover, the ultra-lightweight Conv-GRU CNP module enhances the model's ability to capture spatial information while maintaining a low computational cost. 
The experimental results demonstrate that LSZone excels with a computational complexity of 0.56G MACs and a RTF of 0.37, while delivering impressive performance in noisy and multi-speaker environments, which confirm the potential of LSZone as a highly effective solution for real-time multi-zone speech separation.

\bibliographystyle{IEEEbib}
\bibliography{strings,refs}

\end{document}